# Scalar Diffraction Analysis Of Dispersion In Low-Index Thin Flat Lenses

**ASHFAQUL ANWAR SIRAJI[1,2] AND YANG ZHAO[1,*]**

[1] *Department of Electrical and Computer Engineering, Wayne State University, Detroit, MI, 48202, USA.*
[2] *The MathWorks Inc.*
*Corresponding author: yzhao@wayne.edu*

We analyze the dispersion property of low-index thin lenses by using scalar diffraction and finite difference time domain (FDTD) methods. We compare the dispersion results obtained by using these methods with reported experimental results, and the well-known analytical formula for focal length ($f$) of diffractive lenses as a function of wavelength ($\lambda$), $f(\lambda) = \frac{f_0 \lambda_0}{\lambda}$, where $f_0$ is the designed focal length for wavelength $\lambda_0$. We show that when the analytical formula is applied to thin flat lenses with low-refractive index, the results are accurate for small numerical aperture (NA) up to 0.2. For larger NA, the error between the analytical approximation and the FDTD analysis remains around 8% over a wide range of NA.

Thin-flat lenses are important to the miniaturization of optical devices because of their flat geometry and small size[1–6]. These lenses are more dispersive than traditional lenses and efforts are underway to study the dispersion properties[7–11]. The numerical aperture (NA) of thin flat lenses can be much higher than that of a traditional lens or a diffraction lens. For instance, an NA of 0.4 is considered to be high for a traditional or diffraction lens [12], whereas the NA of a thin flat lens can be close to or even greater than 1 [13]. While lenses with large NA can have novel properties and applications, studying the dispersion of these lenses involves complicated numerical computations [12,14]. The goal of this study is to determine the accuracy of analytical and scalar diffraction methods for the study of dispersion in thin flat lenses with high NA. The study aims at thin flat lenses with low-index materials. Although most thin-lenses use high refractive index materials, low refractive index materials can be an attractive alternative because of their inherently small loss [6,7,15,16].

It has been shown that several approximations of the classical diffraction theory can be used to describe the focal field pattern and dispersion of traditional lenses [12]. These approaches include analytical approximations, scalar diffraction, and vector diffraction theory. They have varying degrees of accuracy and computational complexity. In this work, we employ these methods to study the dispersion of thin flat lenses. We study an analytical approximation derived for diffractive lenses [17,18], a scalar diffraction approximation[6], and a vector finite difference time domain (FDTD) method. We evaluate the accuracy of the scalar diffraction and analytical approximation by comparing these results with available experimental data and FDTD results. We do not include materials dispersion in the study, because most low refractive index materials have low dispersion compared to diffraction dispersion in visible wavelengths.

For diffractive lenses, a well-known analytic approximation for the focal length ($f$) as a function of wavelength ($\lambda$) is [17,18]:

$$f(\lambda) = \frac{f_0 \lambda_0}{\lambda} \qquad (1)$$

where $f_0$ is the designed focal length for wavelength $\lambda_0$. (1) is a good approximation for traditional diffractive lenses with very small NAs. The applicability of (1) to thin flat lenses with higher NA has not been explored. To verify the accuracy of (1), we first calculate the focal length of a thin flat lens at different wavelengths by using the Rayleigh-Sommerfeld (RS) approximation [6] and perform a comparative study with reported experimental results.

The thin flat lenses for this study consist of low-index subwavelength structures [6]. Once the desired focal length of a focusing lens is known, the subwavelength structure of an equivalent flat lens can be designed in various ways. When using low refractive index transparent materials, the subwavelength structure of the flat lens reduces to the filling factor of the material in an air matrix. This is possible because light interacts weakly with low index material, and the phase change is obtained through propagation only. For instance, to focus the beam into a focal spot at a distance $f$, the flat lens must impart the following phase [4,6,10]:

$$\Phi(x,y) = \frac{2\pi}{\lambda}\left(\sqrt{x^2 + y^2 + f^2} - f\right), 0 < \Phi < 2\pi \qquad (2)$$

where $\lambda$ is the operational wavelength and (x,y) are the in-plane spatial co-ordinates. After discretizing the desired phase profile, the required phase at each unit cell (or pixel) of the flat lens is known. This required phase can be implemented by using propagation phase:

$$\Phi_0 = \frac{2\pi n_{eff}(x,y) H}{\lambda} \qquad (3)$$

where $\Phi_0$ is the desired phase at the location (x,y), H is the thickness of the lens, and $n_{eff}(x,y)$ is the required effective refractive index at (x,y).

It is possible to control the effective refractive index of a single pixel of the flat lens by using the filling factor (FF) of the subwavelength structure of the lenses.

$$n_{eff} = \sqrt{n_s^2 FF + 1 - FF} \quad (4)$$

Here, the background material is assumed to be air, $n_s$ is the refractive index of inclusion material ($n_s$ <1.9) and FF is the inclusion filling factor. Each pixel of the lens might be an arrangement of different inclusions with specific filling factor [6]. To implement a specific filling factor using a fixed $n_s$, it is sufficient to implement a filling factor profile FF(x,y):

$$FF(x,y) = \frac{\Phi_{\text{pixel}}^2(x,y) - 2\pi H}{2\pi H(n_s^2 - 1)} \quad (5)$$

Thus, the subwavelength structure of a flat lens that focuses an incoming beam into a focal spot can be designed once the focal lengths is specified.

In the Rayleigh-Sommerfeld (RS) approximation, the complex amplitude of the light at a plane (x, y) which is a distance D away from the lens edge ($x_{in}$, $y_{in}$) along the optical axis can be given as [19]:

$$A(\lambda,x,y) = \frac{D}{j\lambda} \int\int \left(\frac{Te^{-\Phi_{\text{edge}}(x_{in},y_{in})}e^{ikr}}{r}\right) dx_{in} dy_{in} \quad (6)$$

where $\lambda$ is the operating wavelength, T is the transmission of the lens, $\Phi_{\text{edge}}$ is the imparted phase by the lens, $k$ is the wavenumber of the beam, $(x_{in}, y_{in})$ is the coordinate at the lens edge, and r is given as

$$r = \sqrt{D^2 + (x - x_{in})^2 + (y - y_{in})^2}$$

RS approximation (6) was used in [6] to calculate the focal properties of a thin lens. It was shown in [6] that for a single wavelength analysis, predictions made using this equation matches well with vector FDTD simulations and experimental results for NAs up to 0.9.

We compare the results obtained by using the analytical approximation in (1) and the results obtained by using the scalar diffraction in (6) with experimental results from [10]. The results are summarized in Fig. 1 . The lens used in [10] has a numerical aperture of 0.2, a thickness of 600 nm, and a period of 400nm. Though the thin flat lens in [10] do not use low index materials, it relies on geometrical propagation to implement the phase described in (2). The focusing mechanism and the focusing properties of this lens is comparable to the lens we studied in this work.

From Fig. 1, we observe that both the RS approximation and (1) matched the experimental result, with a 2.4% error for the scalar diffraction and a 2.5% error for (1). These results indicate that the scalar diffraction and (1) accurately predicts the dispersion of a thin flat lens if the NA is small.

To study the accuracy of the analytical and scalar diffraction over a large value of NA, we calculated the dispersion of a thin flat lens by using the vector full wave finite different time domain (FDTD) method and perform comparative studies.

Fig. 2 shows the dispersion property of the flat lens with NA = 0.2 and $\lambda_0 = 400nm$, calculated by using the FDTD and the RS methods. It is seen that for NA=0.2, the focal length translates linearly as wavelength increases, and the results obtained using the two methods show agreement. The scalar diffraction theory predicts that the focal length smoothly shifts as the operating wavelength changes, with the shape of the intensity profile preserved, as shown in Fig. 2(b). The vector analysis shows similar behavior. This result indicates that the scalar diffraction and vector FDTD methods agree when NA is small.

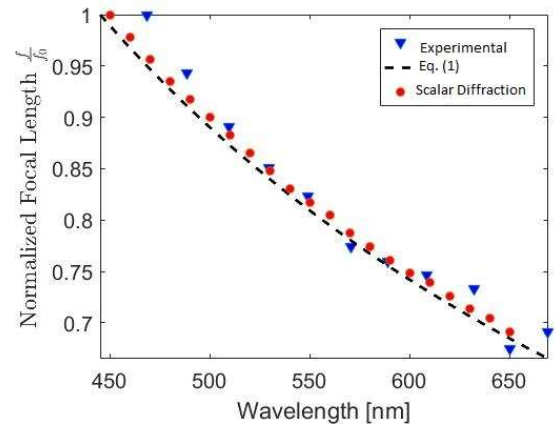

Fig. 1 Comparison of focal lengths shift predicted by using the analytical approximation and the scalar diffraction prediction with experimental results from [10]. The design wavelength ($\lambda_0$) is 530nm, and the NA is 0.2.

We compared the focal length shift for higher NA by using the analytical formula in (1), the scalar diffraction, and the vector full wave FDTD method. The results are summarized in Fig. 3. Results from the scalar diffraction and (1) matches well even at higher NA. For instance, at NA = 0.45, 0.65, 0.85, the scalar diffraction and (1) predict the same shift in focal length, as shown in Fig. 3(a), (b), and (c). The focal length shift obtained by using the vector FDTD method shows certain discrepancy. This is because both (1) and scalar diffraction methods ignore additional effects introduced due to the vector nature and polarizations of light, such as off-axis peaks [12]. In addition, the change in the effective refractive index of neighboring pixels of the thin flat lens is not negligible. Therefore, the scalar diffraction in (6) has limited accuracy compared to an FDTD analysis. These effects lead to a persistent discrepancy between the results from the scalar diffraction in (6) and the vector full-wave FDTD method when studying dispersion.

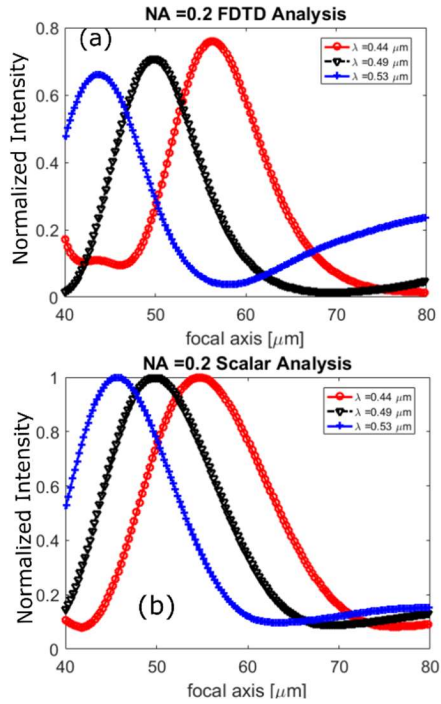

Fig. 2 The light intensity distribution along the optical axis of a thin flat lens calculated by using (a) the FDTD method and (b) the Rayleigh Sommerfeld approximation.

While the scalar diffraction in (6) and the analytical approximation in (1) show deviations from the vector analysis, the error does not increase at higher NA. That is, the scalar diffraction and the approximation in (1) agrees well for a large range of NA. Fig. 4 shows the errors in the normalized focal length shift for the scalar diffraction and (1) when compared to the that calculated by using the full wave FDTD method. For each NA, the error is calculated as:

$$Error(\%) = 100 \times \frac{\left|\left(\frac{f}{f_0}\right)_{FDTD} - \left(\frac{f}{f_0}\right)_{scalar}\right|}{\left(\frac{f}{f_0}\right)_{FDTD}} \qquad (7)$$

The errors of the scalar diffraction and (1) remain around 8% for the NA range 0.3-0.9. These results indicate that (1) is a reasonably accurate method to predict the dispersion of thin flat lenses even at high NA. The scalar RS diffraction method has a similar accuracy.

In conclusion, we have studied the dispersive nature of low-index thin flat lenses and compared three methods for calculating the dispersion. We showed that the analytical dispersion equation (1) originally derived for diffractive lenses remain accurate for low index thin flat lenses with low NA. When the NA is higher, the error between the analytical approximation and the FDTD analysis remains around 8% over a wide range of NA. Because the analytical approximation is significantly faster in computation and easier to use, it can be sufficient for most applications. Full wave FDTD analyses can be used to design and evaluate focal plane light intensity distribution of flat lenses with high accuracy, at the cost of added computational complexity.

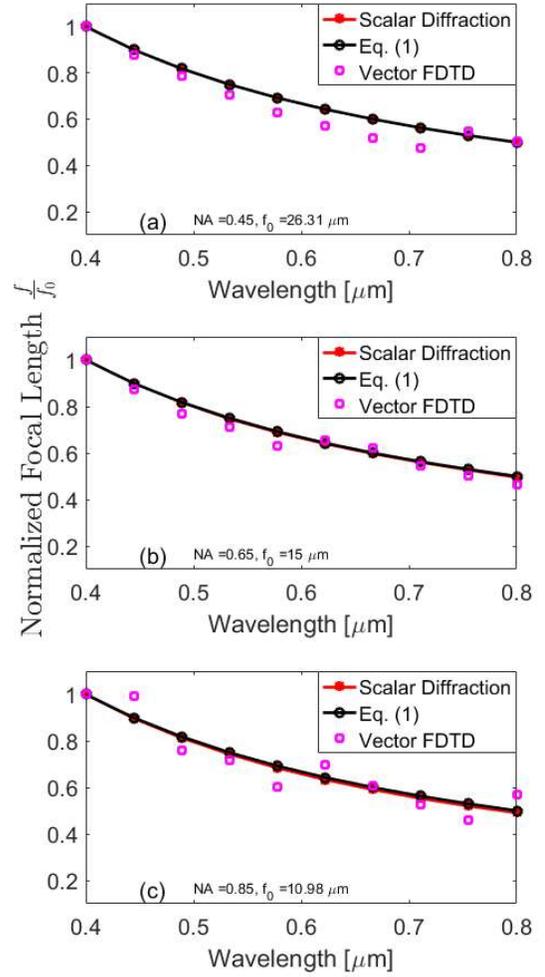

Fig. 3 The dispersion of focal length with respect to operating wavelength of the lens at (a) NA = 0.450, (b) NA = 0.65 and (c) NA = 0.85. The dispersion is calculated by using the scalar RS diffraction equation in (6), the FDTD method, and the analytical approximation in (1).

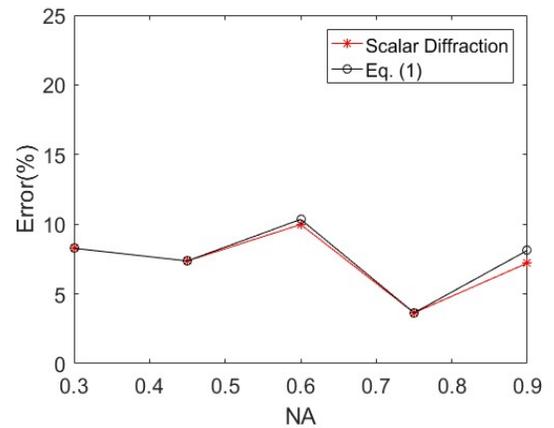

Fig. 4 The error in the scalar and analytical results compared to the FDTD analysis.

**References**
1. N. Yu and F. Capasso, Nature materials **13**, 139 (2014).
2. M. Decker, I. Staude, M. Falkner, J. Dominguez, D. N. Neshev, I. Brener, T. Pertsch, and Y. S. Kivshar, Advanced Optical Materials **3**, 813 (2015).
3. S. J. Byrnes, A. Lenef, F. Aieta, and F. Capasso, Optics express **24**, 5110 (2016).
4. M. Khorasaninejad, W. T. Chen, R. C. Devlin, J. Oh, A. Y. Zhu, and F. Capasso, Science **352**, 1190 (2016).
5. A. She, S. Zhang, S. Shian, D. R. Clarke, and F. Capasso, Optics express **26**, 1573 (2018).
6. A. A. Siraji and Y. Zhao, Applied Optics **58**, 4654 (2019).
7. M. Khorasaninejad, Z. Shi, A. Y. Zhu, W.-T. Chen, V. Sanjeev, A. Zaidi, and F. Capasso, Nano letters **17**, 1819 (2017).
8. S. Shrestha, A. C. Overvig, M. Lu, A. Stein, and N. Yu, Light: Science & Applications **7**, 85 (2018).
9. M. Ye, V. Ray, and Y. S. Yi, IEEE Photonics Technology Letters **30**, 955 (2018).
10. W. T. Chen, A. Y. Zhu, V. Sanjeev, M. Khorasaninejad, Z. Shi, E. Lee, and F. Capasso, Nature Nanotechnology **13**, 220 (2018).
11. L. Li, Q. Yuan, R. Chen, X. Zou, W. Zang, T. Li, G. Zheng, S. Wang, Z. Wang, and S. Zhu, Chin. Opt. Lett. **18**, 082401 (2020).
12. M. Mansuripur, J. Opt. Soc. Am. A **3**, 2086 (1986).
13. M. Khorasaninejad, A. Y. Zhu, C. Roques-Carmes, W. T. Chen, J. Oh, I. Mishra, R. C. Devlin, and F. Capasso, Nano letters **16**, 7229 (2016).
14. Y. Zhang, H. An, D. Zhang, G. Cui, and X. Ruan, Opt. Express **22**, 27425 (2014).
15. A. Zhan, S. Colburn, R. Trivedi, T. K. Fryett, C. M. Dodson, and A. Majumdar, ACS Photonics **3**, 209 (2016).
16. A. A. Siraji and Y. Zhao, Optics letters **40**, 1508 (2015).
17. E. Arbabi, A. Arbabi, S. M. Kamali, Y. Horie, and A. Faraon, Optica **4**, 625 (2017).
18. M. Born and E. Wolf, 7th ed. (Cambridge University Press, 1999).
19. D. G. Voelz, (SPIE press Bellingham, WA, 2011).